\documentclass[letter]{aa} 
\usepackage{multirow}
\usepackage[normalem]{ulem}

\usepackage{graphicx}
\usepackage{txfonts}
\usepackage{footmisc}

\usepackage{capt-of}

\usepackage[colorlinks=true,linkcolor=blue,citecolor=blue,urlcolor=blue]{hyperref}

\begin{document}

\title{Does the solar oxygen abundance change over the solar cycle?}
\subtitle{An investigation into activity-induced variations of the \ion{O}{I} infrared triplet}

 \author{A.G.M.\ Pietrow\inst{1} 
 \and M.\ Baratella\inst{2} 
 \and I.V. Ilyin\inst{1} 
 \and M. Steffen\inst{1}
 \and K. G. Strassmeier\inst{1}
     }

 \institute{\inst{1}Leibniz-Institut für Astrophysik Potsdam (AIP), An der Sternwarte 16, 14482 Potsdam, Germany\\
 \inst{2}ESO – European Southern Observatory, Alonso de Cordova, 3107 Vitacura, Santiago, Chile\\
       \email{apietrow@aip.de}\\
      }

\date{Draft: compiled on \today}

\abstract{The determination of the solar oxygen abundance remains a central problem in astrophysics, as its accuracy is limited not only by models but also by systematics. While many of these factors have been thoroughly characterized, the effect of the solar activity cycle has so far remained unexplored. 
Due to its relative strength and accessibility, the \ion{O}{I} infrared triplet is typically the primary choice for abundance studies. However, previous investigations have shown that abundances inferred from this triplet tend to be higher than expected on active stars, whereas such an overabundance effect is not observed for the much weaker forbidden \ion{O}{I}~6300~\AA\ line.
This raises the question of whether a similar trend can be found for the Sun.
To address this question, we analyze two decades' worth of synoptic disk-integrated Sun-as-a-star datasets from the FEROS, HARPS-N, PEPSI, and NEID spectrographs, focusing on the infrared triplet (7772, 7774, 7775~\AA) and the forbidden \ion{O}{I}~6300~\AA\ line. 
The excellent signal-to-noise ratio of the PEPSI observations allows us to detect a weak but significant variation in the equivalent widths of the infrared triplet, corresponding to about 0.01\,dex difference in abundance between activity minimum and maximum. This value is significantly smaller than the typical uncertainties on the solar oxygen abundance. Due to higher scatter, no comparable trend is found in the other data sets.
Based on these results, we conclude that within the typical uncertainties presented in other works, we can assume the inferred solar oxygen abundance to be stable across the solar cycle, but that this effect may be significant for other, more active stars.}


 \maketitle
%

\section{Introduction}
The solar oxygen abundance\footnote{We adopted the traditional astronomical logarithmic abundance scale $A(\rm O) = 12 + \log_{10}(n_{\rm O} / n_\mathrm{H})$, which expresses the abundance of oxygen on a logarithmic scale relative to $n_\mathrm{H} = 10^{12}$ hydrogen atoms.}, A(O), is a fundamental reference point in astrophysical processes, such as the metallicity of galaxies \citep[e.g. ][]{Zabel2021,LeReste2022}, evolution of stars \citep[e.g.,][]{Vandenberg12}, and the formation properties of exoplanets \citep[e.g.,][]{line21}. Additionally, its abundance strongly affects the opacity of stellar interiors, and thus is an important ingredient in stellar models and evolution codes \citep[e.g.,][]{Basu08, bergemann21}. It is thus crucial to accurately constrain the A(O).

\begin{figure*}
\centering
\includegraphics[width=0.89\textwidth]{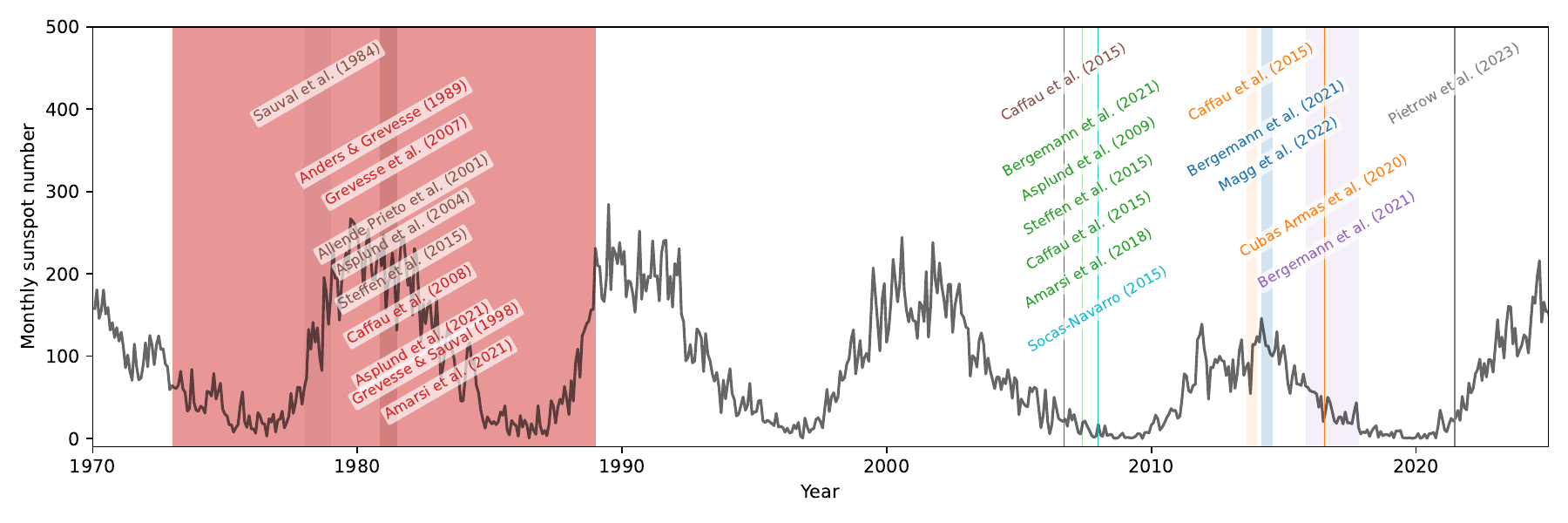}
\caption{Monthly mean sunspot number (black line)) showing the last six solar cycles. Shaded bands indicate the windows used for observing in past solar oxygen abundance studies. These windows are labeled according to their color.}
\label{fig:results}
\end{figure*}

This has been the focus of many studies over the last forty years \citep[][Fig. 7]{Lind2024}. However,  differences are still found when comparing the derived abundance for different data sets. For example several observations from different sources are studied in  \citet{bergemann21} and later \citet{Pietrow2023}. All are processed in the same way, but the inferred values have a scatter of just below 0.1 dex. These discrepancies are generally attributed to instrumental effects such as fringing, continuum placement, and resolution, as well as uncertainties in atomic data including oscillator strengths, line blends, and collisional rates. An additional factor that has not yet been investigated is the possibility of activity-induced spectral line variations over the solar cycle. This is particularly relevant since most A(O) studies have been based on data taken during the solar maximum and the declining phase (see Fig. \ref{fig:results}).

One reason why this potential correlation has not been investigated for oxygen specifically may be the influential study by \citet{Livingston2007}, who showed that weak photospheric lines forming in the
lower photosphere, such as carbon \ion{C}{I}~5380~\AA, do not vary over the solar cycle \citep[e.g., ][]{AlMoulla2022}. In contrast, strong lines forming in the upper photosphere, such as the \ion{Na}{I}~D doublet, exhibited clear correlations with solar activity. 
While it is generally agreed that such differences reflect activity-related biases in the equivalent width (EW) of spectral lines rather than true changes in elemental abundances \citep{2019yana,2020baratella,2020spina}, a definitive explanation is still lacking. Resolving this issue is especially important given that most stellar abundance measurement techniques do not explicitly account for activity effects (although some solutions are being investigated, e.g. \citealt{2024nordlander}).

Nevertheless, the effect of the solar cycle on A(O) has, to our knowledge, not yet been investigated, although several indicators suggest it could be relevant. For example, \citet{Morel2004}
found that abundances obtained from the infrared oxygen triplet 
tended to be higher for more active stars, while such a correlation was not found for the lower-forming forbidden \ion{O}{I}~6300~\AA\ line.  This result was later confirmed by \citet{Schuler2006} and \citet{Shen2007}. This effect was also noticed in galactic abundance studies, although this difference was not attributed to activity but atmospheric parameters \citep[e.g. ][]{Fabbian2009}.

A similar trend has been reported on young and active stars for other elements including for example, \ion{C}{I} and chromium \ion{Cr}{II} \citep{2020baratellagaps}, and most notably barium  \citep[Ba, ][]{Reddy2017, Baratella2021}. Using solar data an attempt to detect changes in photospheric abundances was made by \citet{Pietrow2024}, who analyzed Sun-as-a-star spectra during a strong flare, but no measurable variation was found. However, \citet{Foad2025} reported EW changes in various solar spectral lines between the solar minimum and maximum.  

In this Letter, we utilize four independent long-term synoptic solar datasets to investigate whether the solar oxygen abundance inferred from the \ion{O}{I}~infrared~triplet and the forbidden \ion{O}{I}~6300~\AA\ line exhibit measurable variations over the solar cycle. 

\nocite{armas20, magg2022, caffau15, Socas-Navarro2015, Amarsi18, Steffen15, Asplund2009, Amarsi21, Asplund21, Gervesse98, caffau08, Asplund04, Allende2001, Grevess2007, anders89, Sauval1984}

\begin{figure*}
\centering
\includegraphics[width=1\textwidth]{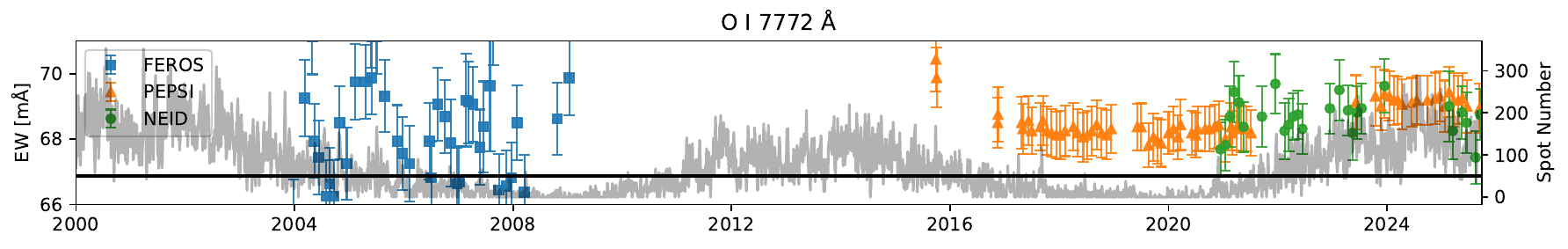}
\caption{Measured EW of the \ion{O}{I}~7772~\AA\ line with FEROS (blue), PEPSI (orange), and NEID (green), including corresponding uncertainties. The gray curve shows the sunspot number as an activity reference, while the black horizontal line at 50 distinguishes active from quiet times. Corresponding plots for the three remaining lines are shown in Fig.~\ref{fig:abundanceVSActivity_full}.
}
\label{fig:abundanceovertime}
\end{figure*}

\begin{figure*}
\centering
\includegraphics[width=0.99\textwidth]{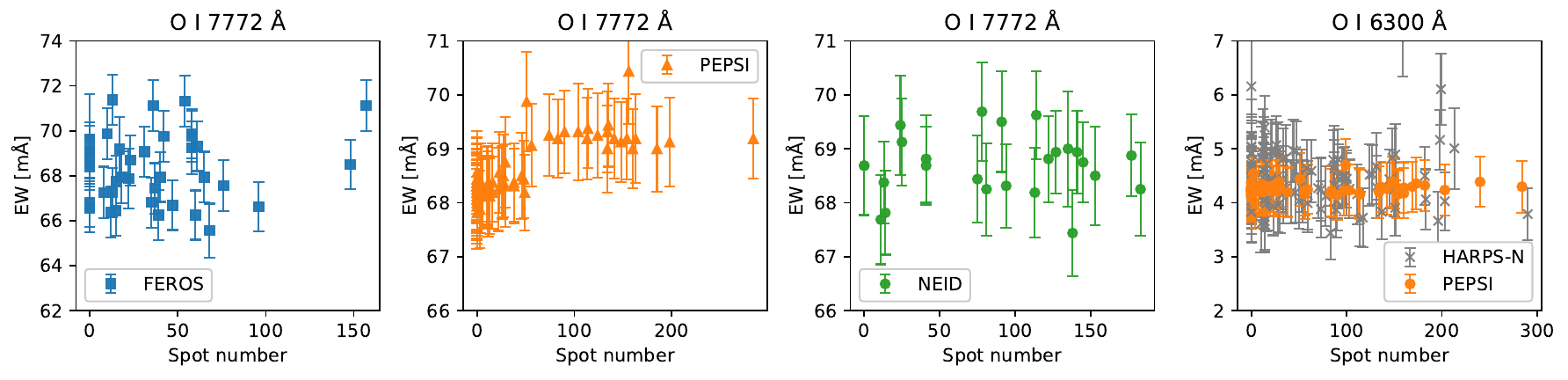}
\caption{EWs of the \ion{O}{I}~7772~\AA\ line for FEROS (blue), PEPSI (orange), and NEID (green) as a function of spot number. Corresponding plots for the three remaining lines are shown in Fig.~\ref{fig:abundanceVSActivity_2}. }
\label{fig:abundanceVSActivity}
\end{figure*}

\section{Observations and data processing}\label{observations}

This study makes use of four Sun-as-a-star datasets, each covering a significant part of a solar cycle. The details and location of each instrument are given below. All signal-to-noise ratios (S/N) are given per pixel and are measured around 5500~\AA.

\subsection{FEROS}
The Fiber-fed Extended Range Optical Spectrograph \citep[FEROS, ][]{Kaufer97} is mounted on the ESO 2.2-m telescope in La Silla observatory in Chile. It has a spectral resolution of \mbox{${\cal R} \approx 48\,000$} and a bandpass between 3500 and 9200~\AA. From November 2003 to May 2011 the instrument had a daily `SolarSpectrum' program where the telescope took scattered-light solar calibration spectra by pointing 30 degrees from the Sun. This was done as a workaround for the halogen calibration lamps which were too weak in the blue. Unfortunately, these observations ceased after the instrument was upgraded with new calibration lamps in 2011 \citep{FEROSmanual}. On average the spectra have an S/N of 200. The data is available on the ESO archive\footnote{\url{https://archive.eso.org/scienceportal/}} under the program id \texttt{60.A-9700}.

\subsection{HARPS-N}
The High Accuracy Radial velocity Planet Searcher in the North \citep[HARPS-N, ][]{Cosentino2012} is a fiber-fed spectrograph mounted on the Galileo National Telescope on La Palma, Spain. Since 2015 it observes the Sun via a dedicated 6-mm solar telescope \citep{Dumusque2015} at a 5-minute cadence. It has a spectral resolution of \mbox{${\cal R} \approx 120\,000$} and simultaneously observes between 3900 and 6900~\AA, at a S/N of 400. Recently, all data from the last decade were released on the DACE platform\footnote{\url{https://dace.unige.ch/sunSearch/?}}  \citep{Dumusque2025-nb}.

\subsection{PEPSI}
The Potsdam Echelle Polarimetric and Spectroscopic Instrument \citep[PEPSI,][]{Strassmeier2015} is a fibre-fed spectrograph installed at the Large Binocular Telescope on Mt.\ Graham, Arizona, USA. It observes the Sun since September 2015 through a dedicated 13-mm solar telescope \citep{Strassmeier2018}. It covers the wavelength range from 3830 to 9070~\AA\, with a spectral resolution of up to \mbox{${\cal R} \approx 250\,000$}, at an S/N of 1000. In July 2023 the solar feed underwent an upgrade to also observe full stokes polarimetry, and resumed operations in June 2024 \citep{Strassmeier2024}. The data is available upon request to the instrument PI.

\subsection{NEID}
The NN-Explore extreme precision Doppler spectrometer \citep[NEID, ][]{Schwab2016} is a fibre-fed spectrograph mounted on the 3.5-m WIYN telescope at Kitt Peak National Observatory, Arizona, USA. It observes the Sun since December 2020 through a dedicated 75-mm solar telescope \citep{Lin2022}. It has a spectral resolution of \mbox{${\cal R} \approx 110\,000$} over a bandpass from 3800 to 9300~\AA, with an average S/N of 320. These data are publicly released through the NEID Solar Archive\footnote{\url{https://neid.ipac.caltech.edu/search_solar.php}}.

\subsection{Data processing}
The data from each telescope were sorted by their S/N ratio, and the highest-quality observations were selected for each month within a three-day window of the 15th, as close to local noon as possible. For FEROS, and NEID a selection was made around the \ion{O}{I}~infrared triplet. For HARPS-N, this wavelength was outside of the instrument's wavelength range and thus the \ion{O}{I}~6300~\AA\ line was considered instead. All four lines were considered for PEPSI.

\begin{table*}

\centering
\small
\caption{Correlation statistics of EWs with solar activity. 
$r$ is the Pearson correlation coefficient, $p$ its significance, 
and the EW medians are given for low ($<$50) and high ($\geq$50) activity levels together with the difference ($\Delta EW$) between the two values. 3D NLTE abundances A(O) are calculated for the two bins, with the matching difference $\Delta A(\rm O)$.}\label{tab:table1}

\begin{tabular}{llccccc|c}
\hline
Inst. & Line & $r$ & $p$-value & EW$_{low}$ [m\AA] & EW$_{high}$ [m\AA] & $\Delta$EW [m\AA]
      & $\Delta A$(O) [dex] \\
\hline

FEROS & \ion{O}{I}~7772~\AA & $+0.139$ & $\phantom{\ll}0.39\phantom{0}$  
      & 68.08 $\pm$ 0.22 & 68.60 $\pm$ 0.33 
      & $+0.52 \pm 0.40$
      & $+0.008 \pm 0.006$ \\

      & \ion{O}{I}~7774~\AA & $+0.19$ & $\phantom{\ll}0.24\phantom{0}$  
      & 59.43 $\pm$ 0.22 & 59.97 $\pm$ 0.33 
      & $+0.54 \pm 0.40$
      & $+0.009 \pm 0.006$ \\

      & \ion{O}{I}~7775~\AA & $+0.385$ & $\phantom{\ll}0.014$ 
      & 46.07 $\pm$ 0.21 & 47.05 $\pm$ 0.33 
      & $+0.98 \pm 0.39$
      & $+0.019 \pm 0.007$ \\
\hline

PEPSI & \ion{O}{I}~7772~\AA & $+0.785$ & $\ll0.001$ 
      & 68.25 $\pm$ 0.11 & 69.26 $\pm$ 0.17 
      & $+1.01 \pm 0.20$
      & $+0.015 \pm 0.004$ \\

      & \ion{O}{I}~7774~\AA & $+0.542$ & $\ll0.001$ 
      & 59.77 $\pm$ 0.09 & 60.55 $\pm$ 0.10 
      & $+0.78 \pm 0.13$
      & $+0.012 \pm 0.002$ \\

      & \ion{O}{I}~7775~\AA & $+0.484$ & $\ll0.001$ 
      & 46.52 $\pm$ 0.09 & 47.04 $\pm$ 0.13 
      & $+0.52 \pm 0.16$
      & $+0.010 \pm 0.003$ \\
\hline

NEID  & \ion{O}{I}~7772~\AA & $+0.046$ & $\phantom{\ll}0.830$  
      & 68.56 $\pm$ 0.27 & 68.70 $\pm$ 0.21 
      & $+0.14 \pm 0.34$
      & $+0.002 \pm 0.005$ \\

      & \ion{O}{I}~7774~\AA & $+0.137$ & $\phantom{\ll}0.51\phantom{0}$ 
      & 59.78 $\pm$ 0.23 & 60.05 $\pm$ 0.17 
      & $+0.27 \pm 0.29$
      & $+0.004 \pm 0.005$ \\

      & \ion{O}{I}~7775~\AA & $+0.253$ & $\phantom{\ll}0.22\phantom{0}$ 
      & 47.09 $\pm$ 0.22 & 47.35 $\pm$ 0.17 
      & $+0.26 \pm 0.28$
      & $+0.005 \pm 0.005$ \\
\hline

HARPS-N & \ion{O}{I}~6300~\AA & $-0.065$ & $\phantom{\ll}0.48\phantom{0}$ 
        & \phantom{0}4.46 $\pm$ 0.07 & \phantom{0}4.31 $\pm$ 0.08 
        & $-0.15 \pm 0.11$
        & -- \\

PEPSI   & \ion{O}{I}~6300~\AA & $+0.165$ & $\phantom{\ll}0.16\phantom{0}$ 
        & \phantom{0}4.24 $\pm$ 0.07 & \phantom{0}4.27 $\pm$ 0.09 
        & $+0.03 \pm 0.11$
        & -- \\
\hline
\end{tabular}

\end{table*}

\section{Methods, results, and discussion}

We used the Automatic Routine for line Equivalent widths in stellar Spectra \citep[ARES v2,][]{2015sousa} code to measure the EWs of the selected lines that automatically fits a Gaussian profile and performs local continuum normalization \citep{2007sousa}\footnote{The EWs reported here are slightly underestimated, as the lines deviate from a purely Gaussian profile and their extended wings are not fully included in the integration (See also footnote~\ref{fnlabel}).}
. The amplitude of these values was then compared with the spot number\footnote{The spot number \citep{Clette2007,Clette2015} is defined as $R = k(10g + s)$, with $g$ the number of groups, $s$ the number of spots, and $k$ a scaling factor depending on the observatory. This metric has been shown to track the \ion{Ca}{II}~H\&K based S-index \citep[e.g. ][]{Bertello2016} despite the Sun being dominated by plages \citep[e.g. ][]{Cretignier20204}.}.  Additionally, two statistical tests are applied to this data for each line separately. First, we calculate the Pearson correlation coefficient (PCC) for EW as a function of spot number. Second, we split the data into two populations and take the median. This results in (1) an active population with spot number $\geq 50$, and (2) a quiet population with spot number $< 50$. 

In Fig. \ref{fig:abundanceovertime} the EWs are plotted over time together with the spot number, which illustrates the solar cycle. In Fig.~\ref{fig:abundanceVSActivity} the EW is plotted against the spot number.

The results of the PCC, together with the median EWs for both groups, are listed in Table~\ref{tab:table1} for each instrument and line. The difference between the inferred EW and abundance values is also given, marked with a delta ($\Delta$), along with the standard uncertainties.

To give a rough indication of the significance of these EW differences between active and quiet, we then computed A(O) for the infrared triplet based on a curve-of-growth equation obtained from the 3D NLTE analysis by \citet{Steffen15}, which are given in Appendix \ref{CoG}. The difference between the high and low abundance values is shown in the same table. No abundance estimate was made for \ion{O}{I}~6300~\AA\ line, as this line is blended with a \ion{Ni}{I} line and requires different modeling.

For the infrared lines, we find that the null hypothesis of the PCC cannot be rejected for either FEROS or NEID, whereas a weak but significant trend appears to be present in the infrared PEPSI EW data. The difference between the high and low median is of the order of 1~m\AA\ for this instrument. When converted into abundances, this corresponds to a change of about 0.01\,dex. This value is below the typical systematic uncertainties of the A(O) computation. 

For the forbidden oxygen line we find no trend in our measurements, which is in line with the findings of \citet{Morel2004}. Using our prior abundance change of 0.01~dex, the expected change in EW would be 0.05 m\AA, which is beyond our precision for this weak line. 

\section{Conclusions}

We investigated the EWs of four synoptic Sun-as-a-star datasets, each covering a significant part of a solar cycle. From these datasets, only the PEPSI data had a high enough S/N ratio to allow the detection of a weak but statistically significant correlation between the solar cycle and the EW of the \ion{O}{I}~infrared~triplet (See table \ref{tab:table1}). 
The PEPSI and HARPS-N observations of the \ion{O}{I}~6300~\AA\ line did not show a trend, likely due to the order of magnitude weaker response of this line to activity, as well as the limited precision of the instruments. 

To illustrate the impact of these changes in EW we converted them into A(O) values based on the 3D NLTE calculations of \citet{Steffen15}, finding a small but consistent change of around 0.01~dex in the PEPSI spectra. This value is well below typical uncertainties of published A(O) values (see Fig. 7 in \citealt{Lind2024}, and the discussions in \citealt{Bergemann2025} and \citealt{Lodders2025}), and thus not relevant for the determination of the solar O(A), but larger activity imprints would amplify this effect as was shown for younger (see the effects on A(O) in the 600 Myr Hyades dwarf stars, \citealt{Schuler2006}), and more active stars (e.g., for $\log R\prime_{HK}>-4.3$ \citet{Morel2004}.

This finding, together with the fact that other stars can exhibit much stronger activity levels, suggests that this could represent a significant source of bias in abundance determinations for younger stars. The high signal-to-noise ratio of the PEPSI observations was crucial for detecting such a subtle trend. Future work will investigate the amplitude of this effect in more active solar analogs and through synthetic Sun-as-a-star spectra with controlled activity filling factors using the Numerical Empirical Sun-as-a-Star Integrator \citep[NESSI, ][]{Pietrow2024nessi}.

\begin{acknowledgements}
AP was supported by grant PI~2102/1-1 from the Deutsche Forschungsgemeinschaft (DFG) and gratefully acknowledges the ESO Chile Scientific Visitor Programme.
\end{acknowledgements}

\bibliographystyle{aa}
\bibliography{ref}

\appendix
\onecolumn 
\section{Curve of growth}\label{CoG}
The following 3D NLTE curves-of-growth are based on the work of \citet{Steffen15}. This study employed a 22-level oxygen model atom, including electron and hydrogen collisions scaled via the SH parameter, which is observationally constrained by the center-to-limb variation of the oxygen triplet. This atom was then used for full 3D non-LTE line formation based on a CO5BOLD
hydrodynamical solar model, NLTE3D for the computation of departure coefficients, and Linfor3D for the generation of synthetic spectra.

These curves were used to convert the measured differences in equivalent width (EW)\footnote{The equivalent width of the oxygen triplet lines given in \citet{Steffen15} are somewhat larger than found in the present work because they were derived from synthetic spectra and include the far wings. \label{fnlabel}} into corresponding differences of the (logarithmic) oxygen abundance A(O) (with $A_0 = 8.76$).:

\begin{align}
\mathrm{EW}_{7772} &= 74.63 
    + 72.43\, (A(\mathrm{O}) - A_0) 
    + 22.11\, (A(\mathrm{O}) - A_0)^2 , \\[4pt]
\mathrm{EW}_{7774} &= 64.31 
    + 65.43\, (A(\mathrm{O}) - A_0) 
    + 20.40\, (A(\mathrm{O}) - A_0)^2 , \\[4pt]
\mathrm{EW}_{7775} &= 50.43 
    + 55.62\, (A(\mathrm{O}) - A_0) 
    + 18.59\, (A(\mathrm{O}) - A_0)^2 .
\end{align}

\section{Further figures}

\begin{center}
  \includegraphics[width=\linewidth]{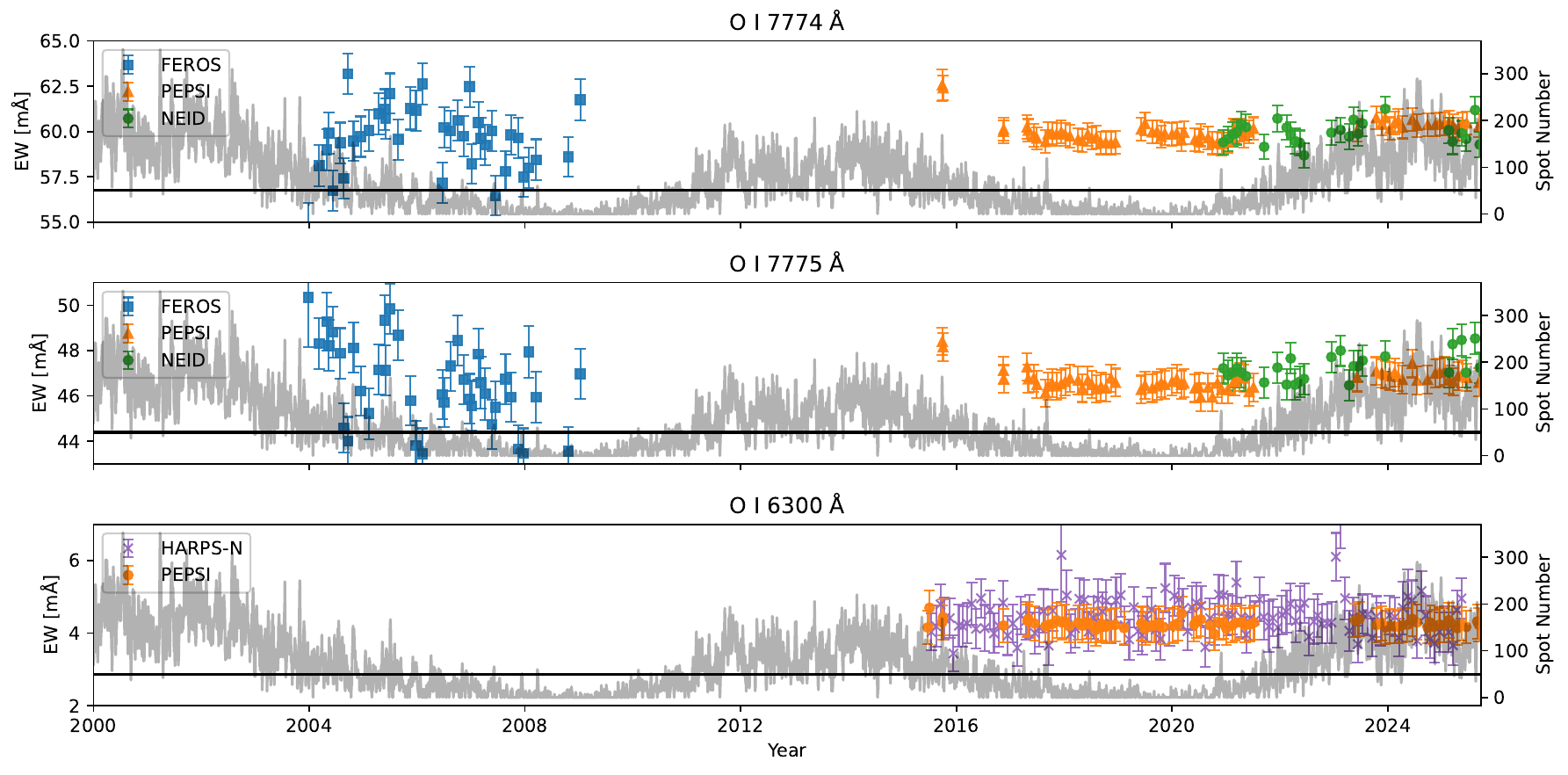}
  \captionof{figure}{Same as Fig.~\ref{fig:abundanceovertime} but for the remaining IR triplet lines and the forbidden line.}
  \label{fig:abundanceVSActivity_full}
\end{center}

\begin{center}
  \includegraphics[width=\linewidth]{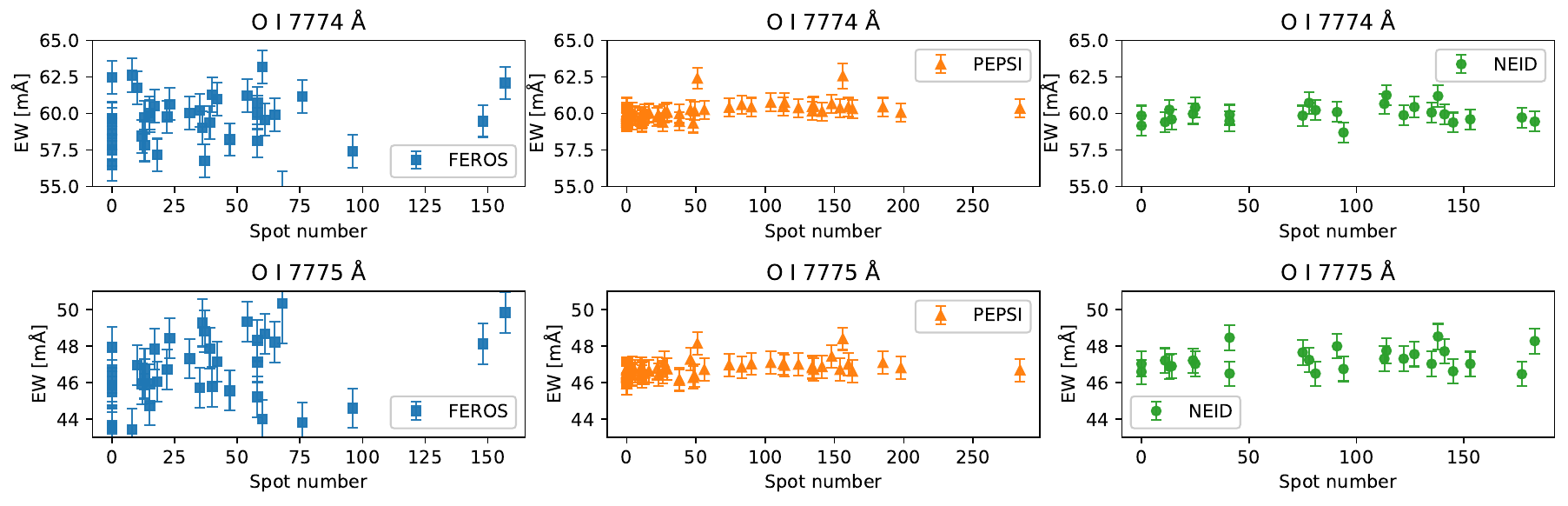}
  \captionof{figure}{Same as Fig.~\ref{fig:abundanceVSActivity} but for the remaining lines.}
  \label{fig:abundanceVSActivity_2}
\end{center}

%
\end{document}